\documentclass[superscriptaddress,pra,showpacs,twocolumn]{revtex4}
\usepackage{graphicx}
\usepackage{epstopdf}
\usepackage{color}
\usepackage{amsmath}
\usepackage{amssymb}

\newcommand{\lpipe}{\rule[-1ex]{0.41pt}{4ex}}

\usepackage{hyperref}
\hypersetup{colorlinks = true, linkcolor = blue}

\begin{document}

\title{Helium diffraction on SiC grown graphene, qualitative and quantitative description with the hard corrugated wall model.}

\author{Maxime Debiossac, Asier Zugarramurdi, Zhao Mu, Petru Lunca-Popa, Andrew J. Mayne and Philippe Roncin}

\affiliation{Institut des Sciences Mol\'{e}culaires d’Orsay (ISMO), CNRS, Univ. Paris Sud, Universit\'{e} Paris Saclay,  Orsay F-91405, France}
\date{\today}

\begin{abstract}
	Monolayer epitaxial graphene grown on 6H-SiC(0001), was recently investigated by grazing incidence fast atom diffraction and analyzed with \textit{ab initio} electronic density calculation and with exact atomic diffraction methods. With these results as a reference, the hard corrugated wall model (HCW) is used as a complementary analytic approach to link binary potentials to the observed atomic corrugation. The main result is that the HCW model reproduces the macroscopic corrugation of the  Moir\'{e} pattern on a quantitative level suggesting that softwall corrections may be neglected for macroscopic superstructures allowing straightforward analysis in terms of a 1D corrugation function.
	 \end{abstract}


\maketitle


\section{Introduction}
Graphene is a promising two dimensional material whose electronic properties depend strongly on its binding to the substrate. This anchoring is visible both in the detailed topography and in the local conductivity, measured in STM experiments where individual contributions are difficult to disentangle. In contrast, AFM or helium diffraction (HAS) are techniques that probe only the topography of the electronic density of the topmost layer\cite{Borca2010,Gibson_2015,Anemone_2016,Maccariello_2016,Taleb_2015,Tamtogl_2015,Taleb_2016,Shichibe_2015}, and are only perturbed by weak polarization effects. This is also the case of grazing incidence fast atom diffraction (GIFAD or FAD) which uses helium atoms with an energy $E_0$ in the keV range but at such a low incidence angle $\theta$,  that the energy $E_\perp = E_0 sin^2\theta$ of the movement normal to the surface is in the sub-eV range. This technique was used recently by Zugarramurdi \textit{et al}\cite{Zugarramurdi_15} together with \textit{ab initio} calculations \cite{Varchon2008} to investigate the structure of an epitaxial graphene mono layer grown on 6H-SiC(0001). Here, we use the much simpler hard corrugated wall model (HCW) to fit the experimental data without \textit{a priori} knowledge of the surface topography nor of the interaction potential. This qualitative description is then evaluated qualitatively by using the HCW with the same interaction potential used in the exact diffraction calculation. The excellent agreement supports the use of the HCW model to investigate the various forms of graphene with fast helium atoms.

\section{GIFAD}
Thermal helium diffraction at crystal surfaces was a seminal experiment of quantum mechanics almost a century ago \cite{Estermann}. The use of fast atoms in the keV range at grazing incidence is much more recent simply because it was not anticipated that atoms with a sub pico-meter wavelength $\lambda$ could be coherently scattered by surface atoms with a thermal fluctuation larger than $\lambda$. Surprisingly GIFAD was discovered, independently in France\cite{brevet,Rousseau2007} and in Germany \cite{Schuller2007}. 
\begin{figure}[h]
	\includegraphics [width=\linewidth]{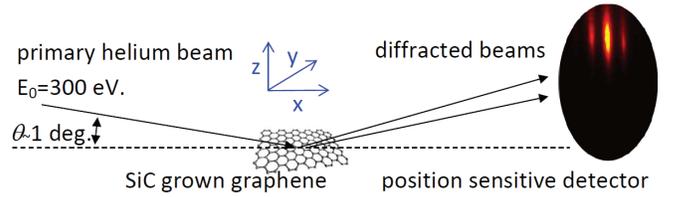} \centering 
	\caption{{Schematic view of a GIFAD setup. A few hundred eV beam of neutral helium atoms impinges at grazing incidence angle onto the surface. When the SiC grown graphene is aligned along a low index direction, diffraction features can be observed onto a position sensitive detector located about one meter downstream.}} 
	\label{GIFAD}
\end{figure} 

It was rapidly noticed that the fast motion along x and the slow motion, in the (y,z) plane, appear decoupled.  In the slow motion plane, the situation is similar to that of a hyper-thermal helium atom, with energy $E_{\perp}$, interacting with a 1D array of quasi-atoms \cite{Danailov}. The fast projectiles are sensitive to the potential averaged along the axial channel, so that only the surface corrugation across the channel is resolved. This forms the basis of the 2D axial surface channeling approximation (ASCA)\cite{Rousseau2007,Schuller2007,Danailov,Zugarramurdi_12,Farias_2004}. The practical interest of GIFAD is that keV atoms are efficiently detected and the full diffraction pattern is 	
confined in a narrow cone which can be recorded in a few seconds using a micro channel plate \cite{Atkinson}.

\section{Theoretical descriptions}
From a theoretical point of view, the scattering of a helium atom with a surface is quite a complex quantum system. First, a potential energy landscape describing the interaction energy $V(x,y,z)$ between the surface and the projectile is required. The transition matrix formalism has been used to identify the different regimes \cite{Manson2008} taking into account phonon excitations. In addition, a purely elastic diffraction regime where the surface atoms are frozen at their equilibrium position providing a periodic 3D potential energy function $V(x,y,z)$ has been investigated using the wave packet technique \cite{Rousseau2007,Aigner2008,Debiossac2014_GaAs} as well as close coupling calculations \cite{Debiossac2014_BSR}, and semi-classical calculations \cite{Gravielle2011,Winter_Review}. The axial surface channeling approximation (ASCA) simplifies the problem, and its range of validity has been investigated in detail \cite{Zugarramurdi_12}. In ASCA, the actual 3D potential is averaged along the low index direction taken here along \textit{x} to generate an effective 2D potential $\tilde{V}(y,z)= \langle V(x,y,z) \rangle _x$. Schematically the "egg box" view of the surface atoms with 2D symmetry is replaced by a "corrugated iron sheet" or washboard description with only 1D symmetry along \textit{y} and translational invariance along $x$. A tiny misalignment of the beam with respect to the low index surface crystallographic axis does not alter the 1D symmetry but simply changes both the effective energy and the initial direction of the effective particle \cite{Seifert_2011,Zugarramurdi_13a,Debiossac_Henkel,Pollak_2015}. With these simplifications, calculating the diffraction of helium atoms onto any periodic topology becomes computationally more affordable.
\section{Graphene preparation} 
Graphene was prepared by epitaxial growth on a highly nitrogen-doped SiC wafer (resistivity: 0.04 $\Omega$.cm, dopant density $3\times 10^{18}$ atoms per cm$^3$). We proceed by starting systematically with the formation of the SiC(0001)-$3\times3$ reconstruction of the Si-face \cite{baffou2008}. Graphene is produced by annealing the Si-terminated surface at 1325$^\circ$C for 25 minutes \cite{Yang2008,Yang2013}. 
The sample was then characterized by STM. Previous observations of layer-by-layer graphene growth on the Si-face of SiC have shown characteristic topographical features due to electronic and geometric contributions that enable the buffer layer, single layer Gr, and bilayer Gr to be unequivocally distinguished in STM \cite{Yang2008,Yang2010,Yang2013,Zugarramurdi_15}. From the STM images, it is estimated that 90\% of the substrate is covered by a single layer. After growth and characterization, the graphene sample was transferred in ambient atmosphere to the GIFAD set-up. A mild annealing at 600$^\circ$C in UHV was sufficient for clear diffraction patterns with a 300 eV He atom beam to be obtained.

\section{Previous results}
In Ref. \cite{Zugarramurdi_15} the reference surface topology was taken from an extensive DFT calculation of the 6H-SiC(0001) surface which includes the intermediate buffer layer and the terminal graphene layer \cite{Varchon2008}. The interaction potential $V(x,y,z)$ with the helium atom was calculated by attaching an effective Lennard-Jones C-He binary potential\cite{Carlos1980} optimized from HAS experiments on graphite considered to be a good model for graphene in terms of the binary potential \cite{Bartolomei2013}.
\begin{figure}[h] \centering
	\includegraphics [width=80mm]{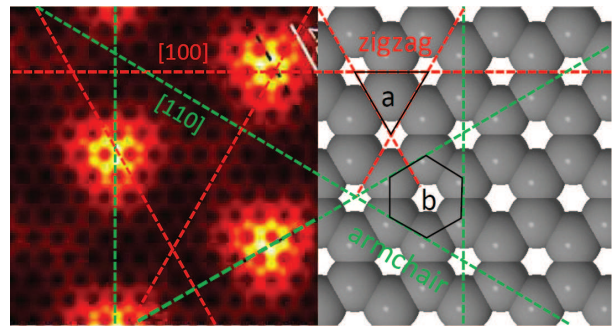} 
	\caption{{left:He-Gr/SiC(001) interaction potential \cite{Zugarramurdi_15} at height $z=2.3\AA{}$. Dashed Color lines indicate the three equivalent zigzag ([100]) and armchair ([110]) directions named according to the graphene honeycomb structure (right). The lattice parameter (a) and the C-C bond length (b) are indicated. The observed Moir\'{e} corrugation is high along the armchair direction connecting neighboring Moir\'{e} bumps and reduced along the zigzag direction connecting a bump to a neighboring valley. Situation is opposite on the graphene backbone where zigzag direction connect holes via the center of C-C bond and have the largest corrugation}} 
	\label{scheme}
\end{figure}

The main results can be summarized as follows. Diffraction was observed along the [100] and [110] directions of graphene corresponding to the zigzag and armchair directions, respectively. The $13\times13R30^\circ$ Moir\'{e} structure of the terminal graphene layer was observed; a precise quantitative measure of the size of the unit cells and relative orientation were obtained and illustrated in Fig. \ref{scheme}. Quantitatively, the LJ potential used to describe the effective C-He interaction was found to be ideally suited in the 10-300 meV energy range where the diffraction on the carbon honeycomb backbone is reproduced without any adjustment. However, the amplitude of the Moir\'{e} structure was overestimated in the calculation \cite{Varchon2008} and the best fit to the data required a scaling factor of 66 \%, i.e. a reduction by 1/3 \cite{Zugarramurdi_15}. 

\section{simplified description within the hard corrugated wall model}
We investigate here the performance of a simpler model, the hard corrugated wall model. The diffraction problem is considered to be a simple reflection of the helium atom at the potential energy line $\tilde{Z}(y)$ corresponding to $\tilde{V_{2D}}(y,\tilde{Z}(y))= E_{\perp}$. This neglects the progressive deceleration/acceleration before and after reflection. The scattering problem is now equivalent to the diffraction of optical rays at a 1D mirror grating described by an arbitrary corrugation function $\tilde{Z}(y)$. In analogy with optics, the diffracted intensities $I_m$ are given by the Fourier transform of the corrugation function $\tilde{Z}(y)$\cite{Garibaldi}; 
\begin{equation}
I_m =\frac{k_{fz}}{k_{iz}}\lpipe\frac{1}{a_y}\int\limits_{0}^{a_y}e^{-imG_y-2i\tilde{k}_\perp\tilde{Z}(y)}dy~\lpipe ^2  
\label{eq:eq1}
\end{equation}
Integration is over one projected lattice unit $a_y=2\pi/G_y$ and  $\tilde{k}_\perp = (k_{iz}+k_{fz})/2$ 
where  $k_{iz}$ and $k_{fz}$ are the initial and final component of the wave vector in the $z$ direction. 
$2\tilde{k}_\perp$ is therefore the momentum transfer in the z direction which, for quasi specular scattering; $mG_y\ll k_{iz}$, is close to $k_\perp\equiv k_{iz}$. At first sight the approximation looks severe, but knowing that most helium-surface interaction potentials have a pronounced exponential decay towards the vacuum, most of the momentum transfer indeed takes place close to the surface of the classical turning point. When the range of the potential and the corrugation amplitude are limited i.e. much smaller than the lattice parameter, this surface is close to the corrugation function $\tilde{Z}(y)$, and the HCW is known to be qualitatively correct\cite{Garibaldi,Armand,Toennies}. However, the accuracy of the HCW model can be limited and corrugation amplitudes with departures up to 30\% from exact theoretical treatments have been reported\cite{Winter_Review}. The main interest of this method is to reduce the complex scattering problem to that of calculating of a 1D integral.
To limit the number of free parameters in the fitting procedures, the corrugation function is expanded, for instance, in a Fourier series $\tilde{Z}(y)=\Sigma_n \alpha_n\cos(nG_{[hjk]}y)$ where $G_{[hjk]}$ is the projected reciprocal lattice vector along the direction $[hjk]$. For a weakly corrugated surface (looking almost flat), the diffracted intensities $I_m$ are usually well described by a single parameter, the full corrugation amplitude $z_c$ corresponding to a single term in the expansion $\tilde{Z}(y)= z_c/2\cos(G_{[hjk]}y)$. The HCW can be solved analytically since it boils down to a very handy Bessel function $I_m=J_m^2(k_{\perp}.z_c)$.  

\section{The zigzag direction}

\begin{figure}[h]\centering
	\includegraphics [width=70mm]{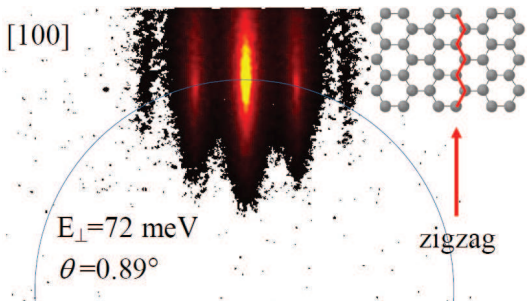}  
	\caption{{Diffraction pattern recorded with 300 eV helium atoms aligned along the graphene [100]  (zigzag) direction. The radius of the Laue circle intercepting the specular spot as well as that of the primary beam, before target insertion, indicates the angle of incidence.}} 
	\label{y13_j067}
\end{figure} 

Along the zigzag direction, Fig. \ref{y13_j067} displays a typical diffraction pattern obtained with 300 eV helium atoms at an angle of incidence of 0.89$^\circ$ corresponding to an energy $E_\perp$=72 meV. The simplest parameter that can be extracted is the peak separation, measured at 0.223$\pm$0.002 deg. i.e. a momentum $G_{[100]}=2.95\pm0.03 \AA{}^{-1}$. This corresponds to a projected periodicity of 2.13 \AA{} matching very well the value of $3b/2$ where b=1.42 \AA{} is the C-C bond length and $a=\sqrt{3}b$ is the 2D lattice parameter of graphene at room temperature (see Fig.\ref{scheme}). 
It is worth noticing that the diffraction pattern exhibits elongated streaks in the vertical direction, rather than small spots of the size of the primary beam. This indicates that the diffraction is not perfectly elastic as observed for instance on surfaces grown by molecular beam epitaxy \cite{Debiossac2014_GaAs}, or on a freshly cleaved crystal \cite{Debiossac2014_BSR,Busch2012}. This is probably due to the presence of defects in the periodic arrangement which limit the coherence length of the surface, and reduces the accuracy of the measurements. 
\begin{figure}[h]\centering
	\includegraphics [width=70mm] {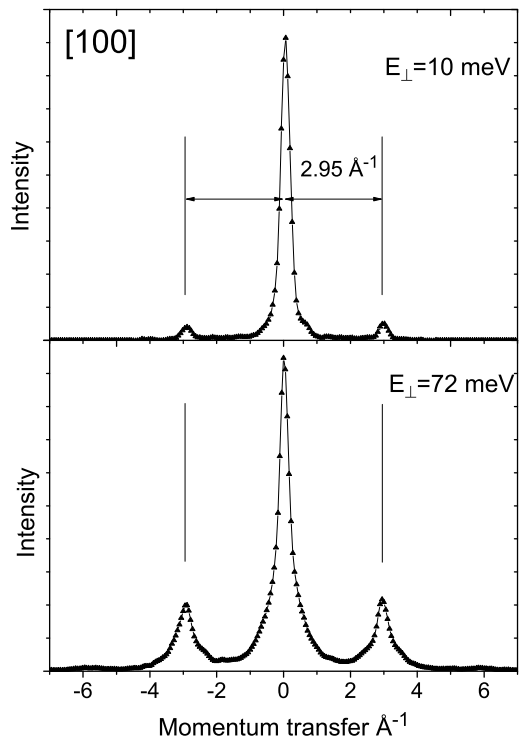} 
	\caption{{For 300 eV helium at 0.33 deg. (a) and 0.89. deg (b) along the [100] direction, the diffracted intensity on the Laue circle (see Fig. \ref{y13_j067}) is reported as a function of the momentum transfer in the perpendicular direction (here [110]).}} 
	\label{y13_j067Px}
\end{figure}

The intensity distribution along the Laue circle is reported in figure \ref{y13_j067Px}. With a perpendicular energy close to 10 meV, mainly specular reflection is observed  and the line profiles are quite narrow (Fig. \ref{y13_j067Px}a). At larger perpendicular energies, figure \ref{y13_j067Px}b shows that the line profile has a more complex structure with shoulders indicating the possible presence of side bands. Despite investigating different regions of the graphene layer, it was not possible to record a diffraction pattern where these satellite peaks could be resolved. We therefore label the diffraction peaks with respect to $G_{[100]}$ associated with the C-C backbone periodicity, and the intensity is considered as that of the whole structure including the contributions of the comparatively large base. 
\begin{figure}[ht]\centering
	\includegraphics [width=70mm] {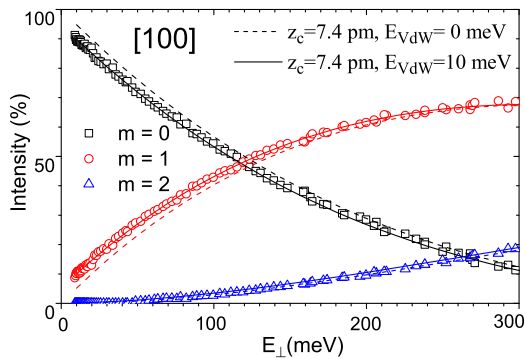} 
	\caption{{Diffracted intensities along the [100] direction as a function of E$_\perp$, the positive and negative diffraction orders have been added. The dashed lines are simple Bessel functions obtained for a sinusoidal HCW of 2.13 \AA{} period and a full corrugation amplitude of 0.074 \AA{}. The solid line shows that the agreement is further improved by adding a $E_{VdW}=10~meV$ energy to $E_\perp$ (see text).}} 
	\label{Im_100}
\end{figure} 
Fig. \ref{Im_100} shows the smooth evolution of the  intensities of the diffraction orders as a function of the angle of incidence expressed here as a perpendicular energy. The dashed lines drawn between the experimental values are the fit of the HCW model for a sinusoidal corrugation function: $I_m=J^2_m(\tilde{k}_\perp z_c)$ 
with a top to bottom amplitude $z_c  = 0.074 \AA{}$. A good agreement with a constant value of the corrugation indicates that the shape of the corrugation function does not change significantly, in this energy range, as the perpendicular energy increases, i.e. as the turning point of the trajectory approaches the surface. A closer look reveals that the dashed lines tend to overestimate the specular intensity at low energy. This could easily be modeled by an increasing corrugation amplitude at low energy. Such a behavior is not uncommon at higher $E_\perp$, where it indicates that the interaction potential is softer on top than on the bottom part of the corrugation function \cite{Momeni2010}. However, such an effect does not seem realistic at all below 30 meV. This behavior is well reproduced here by adding a small Van der Waals contribution $E_{VdW}$ to the effective energy $k_\perp=\sqrt{2M(E_\perp +E_{VdW})}$. This is equivalent to the Beeby correction energy\cite{Sanz2013} used in the Debye-Waller evaluation of thermal decoherence. The value of $E_{VdW} = 10~meV$ for the depth of the Van der Waals potential energy is only an estimation and was not optimized. More accurate values can be derived from bound state resonances \cite{Jardine2004,Debiossac2014_BSR} when the surface coherence is large enough. 
\begin{figure}[ht]\centering
	\includegraphics [width=70mm] {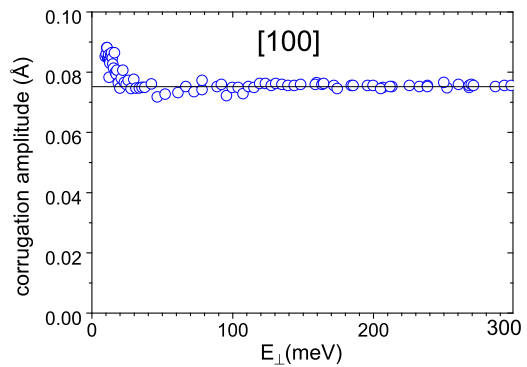} 	
	\caption{{Corrugation amplitude $z_c$ derived from each diffraction image along the [100] direction from a fit of the diffracted intensities $I_m$ by Bessel function $I_m = J^2_m(\tilde{k}_\perp z_c)$ (see text).}} 
	\label{hi_100}
\end{figure}
An other option to retrieve the corrugation amplitude is to fit individual diffraction image with $z_c$ as the only free parameter. Figure \ref{hi_100} shows that $z_c$ is constant above 30 meV giving a value of $z_c  = 0.074 \pm 0.003 \AA{}$. Below 30 meV, the sudden increase of the corrugation amplitude is probably due to the Van der Waals attraction which becomes comparatively important. 

\section{The armchair direction}

\begin{figure}[ht]\centering
	\includegraphics [width=70mm]{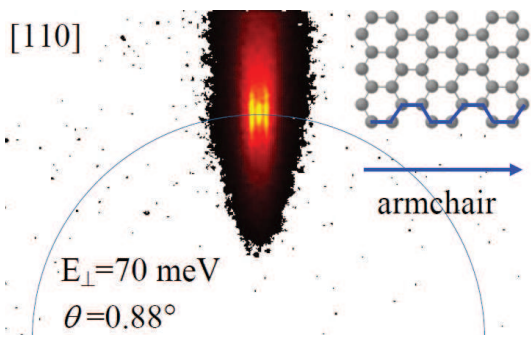} 	
	\caption{{Diffraction pattern recorded along the graphene [110]  (armchair) direction with 300 eV helium at 0.88 deg. incidence corresponding to E$_\perp$ = 70 meV.}} 
	\label{y13_j064}
\end{figure}

\begin{figure}[h]\centering
	\includegraphics [width=80mm]{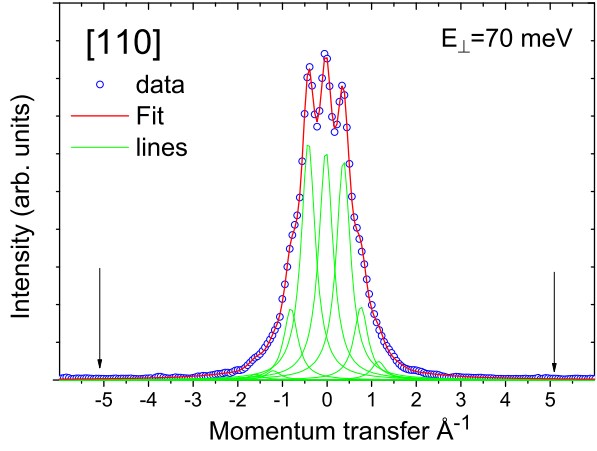} 	
	\caption{{The intensity on the Laue circle of Fig. \ref{y13_j064} is fitted by identical Voigt or Lorentzian profiles having a width $W_L$ and sitting at Bragg position $y_n=nG_{[110]}$. Note the absence of intensity at 5.11\AA{}$^{-1}$ indicated by the vertical arrows. }} 
	\label{Gr_110}
\end{figure}

\begin{figure}[ht]\centering
	\includegraphics [width=80mm]{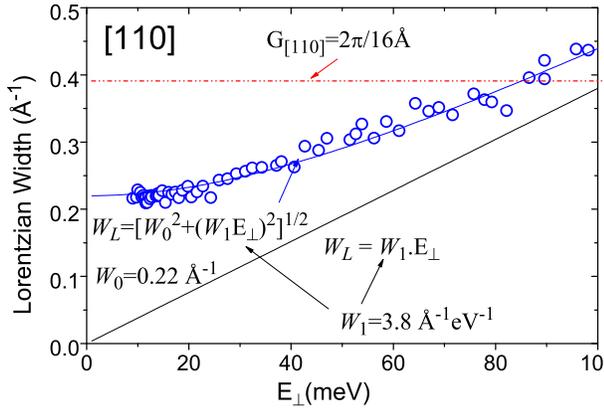} 	
	\caption{{Evolution of the Lorentzian width $W_L$ as a function of the normal energy $E_\perp$ (see text). The straight line indicates a linear dependence with $E_\perp$, while the curved blue line going through data combines quadratically a constant term with this linear one. The horizontal dashed red line indicates the value of $G_{[110]}$. Diffraction signal weakens rapidly when $W_L$ is larger than $G_{[110]}$.}} 
	\label{Wl_110}
\end{figure}

Along the [110] armchair direction, the diffraction pattern is completely different. Fig. \ref{y13_j064} shows a single group of closely packed lines separated by $G_{[110]} = 0.39\pm0.02 \AA^{-1}$. This indicates a projected lattice vector $L_T= 16.1 \pm 0.5$ \AA{} consistent with the Moir\'{e} structure \cite{Zugarramurdi_15}. The counterpart of the C-C backbone structure expected at $5.11\AA{}^{-1} =2\pi/1.23\AA{}$ is conspicuous by its absence (arrows in Fig. \ref{Gr_110}). 
Fig. \ref{Gr_110} displays the diffracted intensity on the Laue circle fitted using Voigt profiles. These are a convolution of a gaussian and a lorentzian, where the gaussian width is that of the primary beam profile $W_G=0.17\AA^{-1}$ while the lorentzian component $W_L$ was left free but is common to all peaks in a given diffraction pattern. Fig. \ref{Wl_110} shows that this width $W_L$ increases gradually with $E_\perp$ and reaches a value larger than the peak separation (the reciprocal lattice vector) when $E_\perp > 90 meV$. Beyond this value the diffraction features are blurred. Defining the visibility $V_{is}$ as the remaining modulation of an infinite array of lorentzians, $V_{is}$ decays exponentially with the ratio of $W_L /G$ but more work is needed to link this visibility to the coherence ratio of the diffracted signal and to the Debye-Waller factor specific to GIFAD \cite{Manson2008,Rousseau2008}. Further studies are also needed to investigate the possible connection between the minimum width $W_0$ and the structural disorder.
 
\begin{figure}[ht]\centering
	\includegraphics [width=80mm]{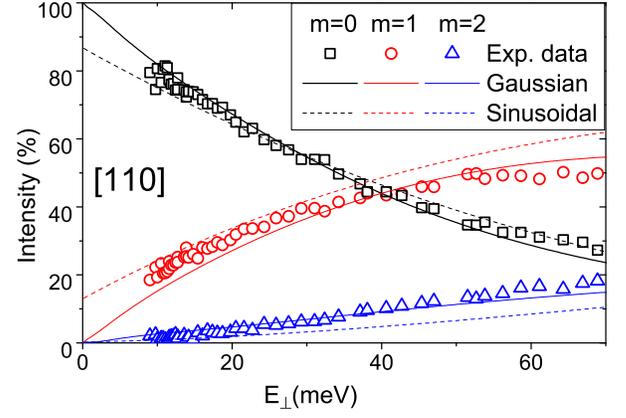} 
	\caption{{Diffracted intensities along the [110] direction as a function of E$_\perp$. The lines correspond to the HCW model with two different functions $\tilde{Z}(y)$ modeling the shape of the Moir\'{e} bumps. Solid lines are for gaussian while dashed ones are for a sinusoidal shape.(see text)}} 
	\label{Im_110}
\end{figure}

Fig. \ref{Im_110} displays the evolution of the five central diffraction orders (m=-2,-1,0,1,2) derived from the fit using Voigt profiles. As in Fig. \ref{Im_100}, the positive and negative diffraction orders have been added together. Note that, due to the deconvolution, the statistical dispersion is significantly larger than along the [100] direction.
Fig. \ref{Im_110} also displays theoretical results using two different corrugation functions for the Moir\'{e} structure, both having the same corrugation amplitude of $z_c\sim0.14\AA{}$.
 Compared with the measured lattice parameter $L_T = 16 \AA{}$, this corrugation amplitude is only 1 \%, which underlines the high sensitivity of GIFAD. The dashed lines are produced by a sinusoidal shape $\tilde{Z}(y)=z_c/2\cos(G_{[110]}y)$ where the valleys and bumps are identical. Fig. \ref{Im_110} indicates that this sinusoidal shape tends to underestimate the population of large diffraction orders, i.e. larger deflection angles, suggesting that steeper slopes are present (larger derivative of $\tilde{Z}(y)$). This can be provided by adding higher order terms in the harmonic expansion of $\tilde{Z}(y)$, but we have chosen to use a gaussian shape 
$\tilde{Z}(y) =z_c e^{-{(y/2L_T\sigma)}^2}$ were only one additional parameter is involved, namely the width $\sigma/L_T$ relative to the lattice parameter $L_T = 16 \AA{}$. The result reported in Fig. \ref{Im_110} is for a relative fwhm of 1/3. Note that, for symmetric profiles, i.e. if a value $y_s$ such that $\tilde{Z}(y-y_s)=\tilde{Z}(y_s-y)$ exists, the HCW model is not sensitive to the sign of the corrugation function $\tilde{Z}(y)$ because it displays the same corrugation amplitude and the same slope distribution. This means that bumps with a fwhm of 2/3 and narrower valleys of 1/3 would produce the same results. These simple models are only indications, it is not completely meaningful to compare the present data where only a few diffraction orders are observed with a complex corrugation function $\tilde{Z}(y)$. 
As a rule of thumb, the number of significant points in the unit cell should compare with the number of observed diffraction orders.
For larger corrugation amplitudes as measured along the $[1\bar{1}0]$ direction of the $\beta_2$(2$\times$4) reconstruction of GaAs(001), up to a hundred diffraction orders have been recorded and much more complex corrugation profiles have been derived \cite{Debiossac2014_GaAs}.

\section{Discussion} 
The experiment provides model independent information such as the peak positions, peak profile, width and intensities. Conversely, the corrugation amplitudes reported here are derived through the HCW model. On the one hand the HCW model can be seen as a powerful tool to describe, with only one or two parameters, the rapidly varying set of data presented in Figs. \ref{Im_100} and \ref{Gr_110}. On the other hand the question remains how accurate is this description? 
It is important to note that the observed intensities result from interferences between trajectories bouncing along the top or bottom of $\tilde{Z}(y)$. Assuming a corrugation amplitude varying slowly with the energy $E_\perp$ the intensity of the specular peak undergoes a full oscillation from dark to bright and dark again in a wavenumber interval $\delta k_\perp$ given by $\delta k_\perp.z_c=1$. i.e. $\delta k_\perp = z_c^{-1}$. This offers a robust and redundant evaluation of the corrugation amplitude $z_c$. However, the HCW model is known to be only qualitative when softwall corrections are needed, i.e. when the sudden momentum transfer approximation is not valid. In this case the classical turning points are no longer close to the equipotential line. The other well-known situation where the HCW fails to be quantitative occurs when the corrugation amplitude becomes large compared to the size of the lattice unit cell giving rise to possible multiple scattering effects. Since an exact diffraction calculation has been performed, it is easy to check if the corrugation amplitude extracted from the data is quantitative or not by applying the HCW model to the same potential energy landscape. It is straightforward to reconstruct the surface potential from the effective Lennard-Jones (LJ) potentials \cite{Carlos1980} used in \cite{Zugarramurdi_15}. However, as shown in Fig.\ref{y13_j067} and \ref{y13_j067Px} and in ref.\cite{Zugarramurdi_15}, the diffraction along the [100] direction is not directly affected by the Moir\'{e} superstructure, thus, along this direction the graphene can be modeled as a flat layer. The pair-wise potential used in the present study is given by
\[V_{LJ}(R)=4\epsilon\left(\frac{\sigma^{12}}{R^{12}}-\frac{\sigma^6}{R^6}\right)\]
with $\epsilon=1.4 meV$, $\sigma=2.74\AA$\cite{Carlos1980}. We use now the string approach where averaging is calculated separately for each row of perfectly aligned atoms with the associated cylindrical coordinate $\rho=\sqrt{y^2+z^2}$ and $n=1/d_x$ the surface atom density along $x$. The cylindrical potential for each string is then given by:
\[V_{LJ}(\rho)=n\pi\epsilon\left(\frac{63~\sigma^{12}}{64~\rho^{11}}-\frac{3~\sigma^6}{2~\rho^5}\right)\]
For both directions investigated here, only two strings of carbon atoms having equal linear densities are present inside the projected lattice unit. Along the [100] zigzag direction, the two rows group together with a separation of $0.71\AA{}$ while the next nearest row, in the next lattice unit, is twice as far away at $1.42\AA{}$. This produces the observed corrugation where the highest points correspond to the middle of these double rows and the lowest points are sitting in between them. Along the  [110] armchair direction, the two rows are evenly spaced minimizing the corrugation amplitude which is indeed almost ten times less than along the [100] direction. The highest points of the corrugation function correspond here to the top of the row, while the lowest points are in between. 
\begin{figure}[ht]	\centering
	\includegraphics[width=\linewidth]{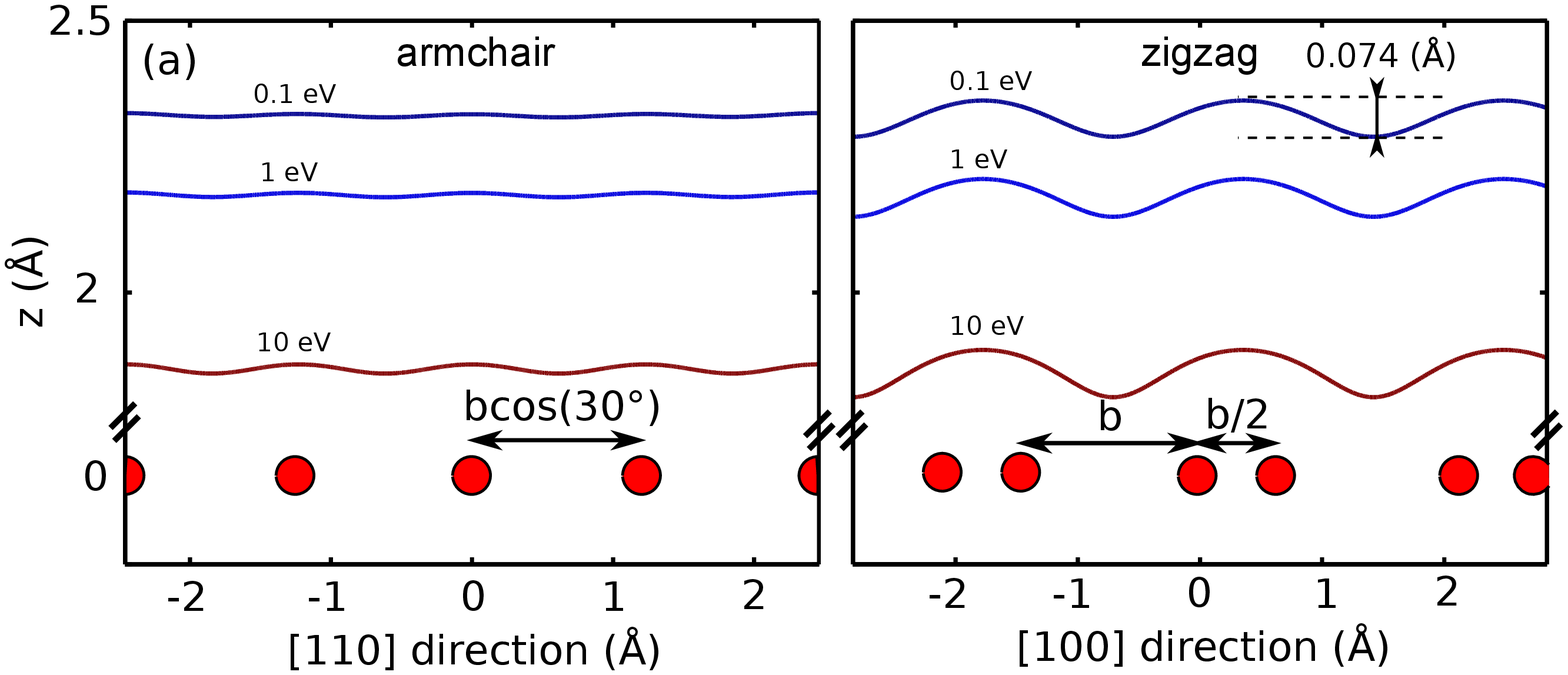} 
	\caption{{Equipotential energy lines of a flat graphene, note that the coordinates are perpendicular to the beam direction. At $E_\perp=100 meV$ a full corrugation amplitude of 0.074 \AA{} is measured. The lowest point in energy is sitting in the middle of the C-C bond (b).}}
	\label{potentiels_[100]}
\end{figure}

\begin{figure}[h]	\centering
	\includegraphics[width=70 mm]{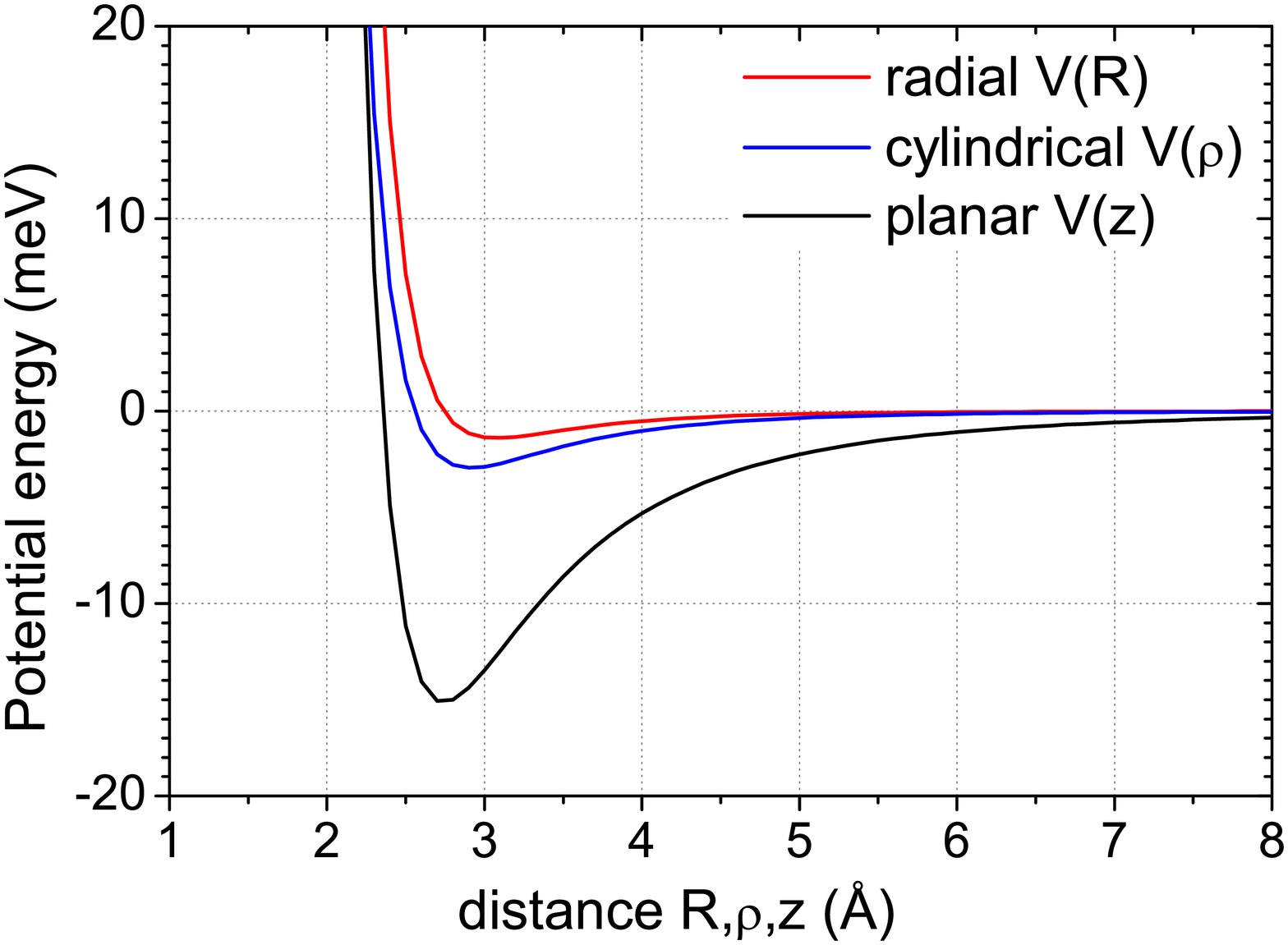}
	\caption{{The Lennard-Jones effective C-He potential\cite{Carlos1980} $V(R)$ is reported together with its value integrated axially $V(\rho)=\int_x V(R)$ and its planar average $V(z)=\int_x\int_y V(R)$. }}
	\label{polarisation}
\end{figure}

The projected potential energy landscape is then easily calculated by adding the contributions of a few lattice units. The equipotential energy lines are calculated numerically and are displayed in Fig. \ref{potentiels_[100]}. The corrugation amplitude calculated at $E_\perp= 100 meV$ is $z_c=0.074\AA{}$, exactly the one derived from the experimental data. This $0.001\AA{}$ accuracy is accidental since this LJ binary potential predicts that $z_c$ should vary from $0.07\AA{}$ to $0.08\AA{}$ between $E_\perp=50 meV$ and $E_\perp=300~meV$ which is not observed in the data. This indicates that the stiffness of the LJ potential may not be entirely correct. This is not too surprising since the LJ form of the repulsive part is purely empirical, while the prefactor has been adjusted  to fit the bound states inside the potential well (see Fig. \ref{polarisation}). 

The other indication provided by the LJ potential is the depth of the Van der Waals potential energy well. This is easily visualized in the planar form obtained by integration along both the $x$ and $y$ coordinates
\[V_{LJ}(z)=2a_s\pi\epsilon\left(\frac{2\sigma^{12}}{5z^{10}}-\frac{\sigma^6}{z^4}\right)\]
where $a_s=2/5.24\AA{}^2$ is the surface density, i.e. the number of atoms in the unit cell divided by its area. The attractive parts of neighboring atoms add up significantly because the location of the well is at a distance larger than the lattice parameter. In contrast Fig.\ref*{potentiels_[100]} shows that the repulsive part of the planar, axial and radial potentials are not too different in the energy range of $E_\perp \sim 10-100~meV$ probed here. The resulting atomic radius, string radius and minimum distance of approach to the surface defined as the distance where the potential energy equals the initial kinetic energy are close to each other.

Along the armchair [110] direction, the graphene carbon backbone is not observed. This is only due to the averaging which reduces the 2D egg-box corrugation from 0.2\AA{} corresponding to center of the hexagon and top of a C atom down to an effective projected corrugation of only 0.007\AA{}. Consequently the graphene backbone is very flat when observed along the [110] direction. The same averaging effect also affects the Moir\'{e} structure but along the zigzag direction because the Moir\'{e} domes are aligned along the [110] as stated by the R30$^\circ$ index.
The corrugation amplitude of 0.14\AA{} derived from the data in Fig. \ref*{Im_110} is in accordance with the value of 0.21\AA{} plotted in Fig. 4c of Ref.\cite{Zugarramurdi_15} before multiplying by the scaling factor of 0.656 resulting in a value of 0.14\AA{}.
The main result here is that fast atom diffraction on Graphene is almost a perfect system to apply the HCW approach.

This conclusion might be less valid for HAS because a 3D potential $V(x,y,z)$ must be considered so that the minimum of the corrugation function is in the center of the hexagon and is not weakened by any averaging. At $E_\perp=100~meV$, the 3D corrugation amplitude is almost 3 times larger than the 2D average along the [100] direction $\tilde{V}(y,z)$. This comparatively strong attenuation is specific to the hexagonal symmetry, and is even more pronounced along the [110] direction. For rectangular lattices there is often a direction where the measured corrugation amplitude compares with its 3D value derived from HAS.

\section{HCW in GIFAD and HAS} 
The corrugation amplitude of the  Moir\'{e} pattern of graphene deposited on various substrates has been investigated by thermal energy helium diffraction (HAS). Corrugations of 0.06$\AA$ on Ni(111) \cite{Tamtogl_2015}, 0.15$\AA$ on Ru(0001) and up to 0.9$\AA$ on Rh(111) \cite{Gibson_2015} have been observed indicating variations of the graphene interaction with the underlying metal substrate. The value measured here by GIFAD on 6H-SiC(0001) is 0.14$\AA$, corresponding to a 2D corrugation of 0.27$\AA$ \cite{Zugarramurdi_15} where the difference between these two values originates from the averaging of the surface potential (ASCA) due to the grazing incidence geometry. The value appears rather large for a system assumed to be weakly bound to its substrate but here the observed corrugation is expected to reflect mainly the corrugation of the underlying buffer carbon layer which is indeed strongly bound to the SiC substrate.

It should be noted that Eq.\ref{eq:eq1} is valid only for normal incidence of the projectile. In GIFAD this corresponds to a primary beam well aligned with the probed direction\cite{Winter_Review,Zugarramurdi_13a}, whereas in a typical HAS setup this orthogonality of both the beam and detector is difficult to achieve. As a result the intensities observed in HAS for opposite diffraction orders, i.e. $+m$ and $-m$, differ limiting the interest of the HCW model in data analysis (see e.g. Fig.3 of \cite{Gibson_2015}). Recently, two independent papers\cite{Debiossac_Henkel,Pollak_2015} have proposed analytical formulae taking into account the oblique incidence. These predictions have not yet been compared together but both should significantly improve the comparison of HAS data with simple descriptions of the surface corrugation function. 

A more significant difference between GIFAD and HAS lies in the ability of HAS to measure phonon modes of the graphene surface by inelastic scattering \cite{Maccariello_2016,Taleb_2015,Tamtogl_2015,Taleb_2016,Shichibe_2015}. 
In the present case some modes specific to quasi-free standing graphene were identified \cite{Maccariello_2016} providing a very interesting link with dynamical behavior of the graphene layer as probed by Atomic Force Microscopy \cite{Koch_2013}.
It would certainly be worth trying to obtain comparable information from scattering of fast atoms. On the one hand, a few meV resolution on top of few hundred eV projectiles does not seem realistic. On the other hand, Shichibe \textit{et al}.\cite{Shichibe_2015} have shown that this inelastic behavior also has specific signatures in the scattering profile perpendicular to the surface plane. Provided that topological defects are under control, these aspects could probably be investigated at grazing incidences as well. 

\section{Conclusion} 
The hard corrugated wall model was found to be fully quantitative with predictions almost indistinguishable from that of an exact close coupling calculation. The reason for the very good agreement lies in the compact interaction potential between the helium atom and the carbon atoms of the surface. In other words, the so called soft potential effects, which tend to separate the turning point surface from the iso-energy surface are probably very small. 
The other favorable condition is that the specific averaging associated with the hexagonal symmetry, generates a 1D apparent corrugation that is significantly weaker that the 2D corrugation. As a result, for $E_\perp$ below 100 meV, the HCW model can safely be used to analyze the numerous Moir\'{e} structures of graphene, and as demonstrated in \cite{Zugarramurdi_15} that the LJ effective binary potential \cite{Carlos1980} can also be useful to generate a potential energy landscape from the atomic positions. For flat graphene, this landscape is derived analytically from the LJ effective binary potential. Furthermore, along the armchair direction, only the overall periodic shape of the Moir\'{e} structure is needed since the individual carbon atoms of the graphene backbone do not contribute to the diffraction pattern.

\section{Acknowledgment}
We are most grateful to Andrei Borisov for continuous discussions and fruitful advice. This research was supported by Triangle de la physique (2012$-$040T$-$GIFAD) and by the Agence Nationale de la Recherche ANR$-$11$-$EMMA$-$0003.

\end{document}